\documentclass[%
    reprint,
    nofootinbib,
    amsmath,amssymb,
    aps,
    prd,
    floatfix,
]{revtex4-2}

\usepackage{graphicx}
\usepackage{dcolumn}
\usepackage{bm}
\usepackage{hyperref}

\usepackage{color}
\usepackage{enumitem}

\begin{document}

\preprint{APS/123-QED}

\title{The muon Moonshot: Moon subsurface tomography with upward-going muons}

\author{Zimo Hu}
\email{zmhu25@stu.pku.edu.cn}
\affiliation{School of Physics and State Key Laboratory of Nuclear Physics and Technology, Peking University, Beijing 100871, China}

\author{Leyun Gao}
\email{seeson@pku.edu.cn}
\affiliation{School of Physics and State Key Laboratory of Nuclear Physics and Technology, Peking University, Beijing 100871, China}

\author{Zhengyun You}
\affiliation{School of Physics, Sun Yat-sen University, Guangzhou 510275, China}

\author{Qite Li}
\affiliation{School of Physics and State Key Laboratory of Nuclear Physics and Technology, Peking University, Beijing 100871, China}

\author{Qiang Li}
\email{qliphy0@pku.edu.cn}
\affiliation{School of Physics and State Key Laboratory of Nuclear Physics and Technology, Peking University, Beijing 100871, China}

\author{Yuhong Yu}
\affiliation{School of Nuclear Science and Technology, University of Chinese Academy of Sciences, Beijing 100049, China\\Guangdong Laboratory for Advanced Energy Science and Technology, Huizhou, Guangdong 516000, China\\Institute of Modern Physics, Chinese Academy of Sciences, Lanzhou 730000, China}

\author{Liangwen Chen}
\affiliation{School of Nuclear Science and Technology, University of Chinese Academy of Sciences, Beijing 100049, China\\Guangdong Laboratory for Advanced Energy Science and Technology, Huizhou, Guangdong 516000, China\\Institute of Modern Physics, Chinese Academy of Sciences, Lanzhou 730000, China}

\author{Xueheng Zhang}
\affiliation{School of Nuclear Science and Technology, University of Chinese Academy of Sciences, Beijing 100049, China\\Guangdong Laboratory for Advanced Energy Science and Technology, Huizhou, Guangdong 516000, China\\Institute of Modern Physics, Chinese Academy of Sciences, Lanzhou 730000, China}

\author{Zhiyu Sun}
\affiliation{School of Nuclear Science and Technology, University of Chinese Academy of Sciences, Beijing 100049, China\\Guangdong Laboratory for Advanced Energy Science and Technology, Huizhou, Guangdong 516000, China\\Institute of Modern Physics, Chinese Academy of Sciences, Lanzhou 730000, China}

\collaboration{PKMu Collaboration}
\noaffiliation


\begin{abstract}
We propose a novel muon Moonshot concept for lunar subsurface tomography based on upward-going muons originating from the lunar regolith. Unlike Earth, the Moon lacks an atmosphere, leaving dense regolith below and a near-vacuum environment above. Consequently, most downward-going hadrons are absorbed before decaying, while upward-going hadrons escaping the regolith can decay in flight, producing a significant flux of lunar muons. These muons can be detected by instruments on the lunar surface or in near-lunar orbit. We perform Monte Carlo simulations to investigate their energy spectra, angular distributions, and integrated fluxes under various theoretical and detector configurations. The results indicate that the lunar muon flux is sensitive to detector altitude under idealized geometry. This dependence demonstrates its potential as a novel, non-invasive probe of shallow subsurface voids. We also present case studies on the detection of underground cavities and water resources. For shallow overburden, large cavity signals remain detectable with estimated backgrounds and systematics. Weaker water signals remain challenging. Potential implementations in future lunar missions are also discussed.
\end{abstract}

\maketitle


\section{Introduction}


Cosmic-ray muons produced in Earth's atmosphere are highly penetrating particles with widespread scientific and industrial applications. Muon tomography~\cite{Bonechi:2019ckl,Checchia:2016vnv} provides a powerful tool for non-invasive imaging of large-scale structures, including volcanoes~\cite{Marteau:2012zv,cheng2022imaging}, pyramids~\cite{Morishima:2017ghw,Procureur:2023wrn}, underground tunnels~\cite{Guardincerri:2017qlx,Thompson:2019bcl}, and nuclear reactors~\cite{Perry:2013bja,Fujii:2019kwi}. By contrast, although the Moon, Earth's only natural satellite, is exposed to a comparable primary cosmic-ray flux that also generates abundant secondary muons, its secondary-particle environment is markedly different due to the absence of an atmosphere.

Muon production and propagation within lunar regolith have attracted interest for underground positioning~\cite{leone2025positioning,prettyman2019deep,enqvist2021exploration,kuusiniemi2023muon}. However, the high density of the lunar regolith results in a short interaction length for meson--nucleus collisions, causing most mesons to undergo repeated interactions before decaying. In contrast, the near-vacuum environment above the lunar surface allows mesons escaping the regolith to decay in flight, producing a significant upward-going lunar muon flux. This source has largely been overlooked, although backscattered muons have recently begun to attract attention~\cite{pinto2023mapping}. These upward-going muons can retain information about the shallow subsurface structure of the regolith, a possibility that has yet to be systematically explored.

Unlike lunar muons, other lunar secondary radiation has been extensively investigated. For example, lunar X-ray emission has been studied through orbital X-ray fluorescence measurements~\cite{adler1972apollo15,adler1972apollo16,gloudemans2021re,swinyard2009x,okada2009x}, in which secondary photons generated within the lunar surface by incident solar radiation are used to infer its elemental composition. These studies demonstrate that secondary radiation escaping the lunar surface carries information about its bulk composition and large-scale geological structure. Motivated by this analogy, secondary cosmic-ray muons may likewise provide a complementary probe of the density and composition of the lunar near surface.

In addition, lunar gamma rays have been studied through seven years of observations by the \textit{Fermi} satellite~\cite{Fermi-LAT:2016tkg,Loparco:2017isq}. These observations motivated the development of a comprehensive Monte Carlo (MC) framework incorporating primary cosmic-ray protons and helium nuclei, together with solar modulation effects. Lunar gamma-ray fluxes have also been proposed as probes of dark matter scenarios~\cite{Garani:2019rcb}, while discrete gamma-ray lines have been investigated as tracers of cosmic-ray activity over different timescales~\cite{fujiwara2025moon}. These gamma-ray studies provide benchmarks for our modeling framework. In this work, we investigate the escaping lunar muon flux as a complementary probe of near-surface structure. We simulate the energy and angular distributions of the upward-going particles to assess the potential for subsurface tomography using surface and orbital detectors.


\section{Event simulation}

\begin{figure}
\includegraphics[width=\columnwidth]{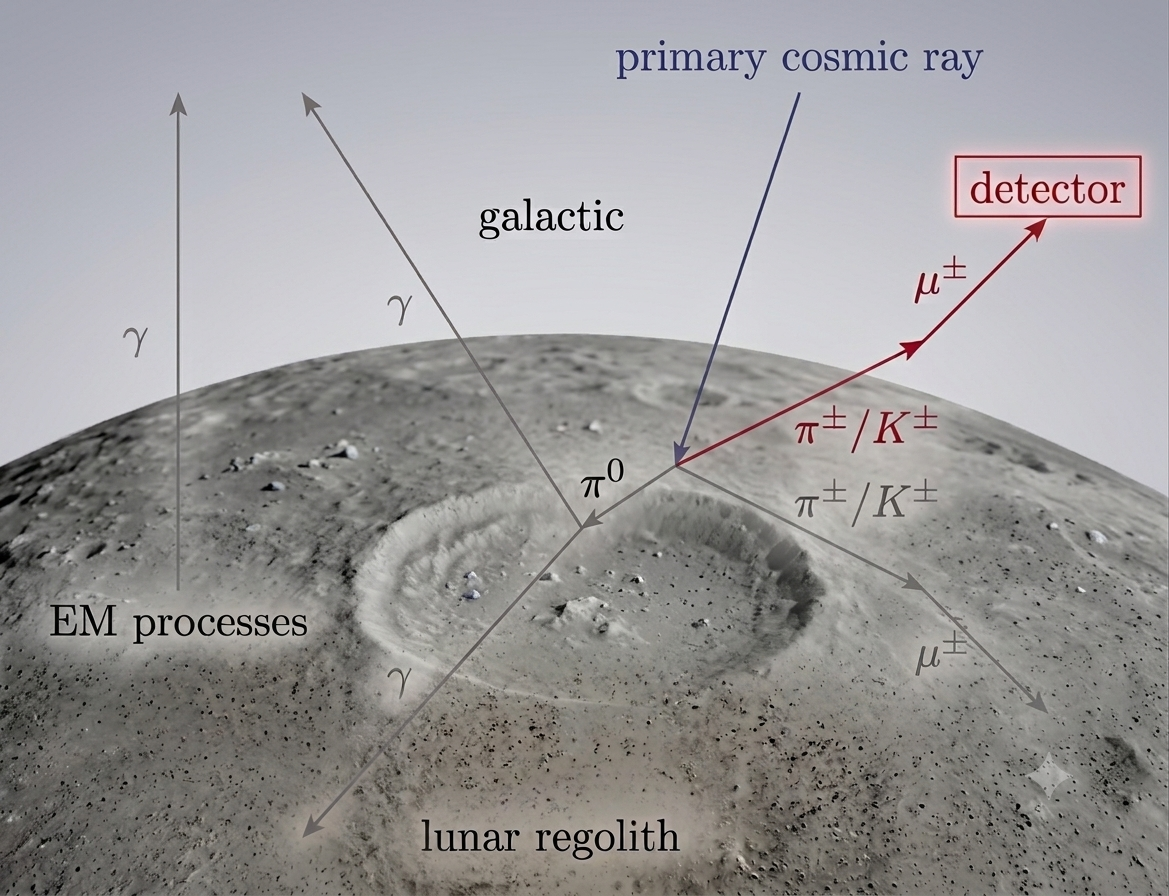}
\caption{Schematic illustration of the production of mesons, muons, and photons resulting from the interaction of primary cosmic rays with the lunar regolith. Muons escaping from the lunar surface can be detected by instruments located on the surface or at a height, providing a means for lunar subsurface tomography.}
\label{fig:muon-moon-shot}
\end{figure}

As illustrated in Fig.~\ref{fig:muon-moon-shot}, cosmic-ray interactions with the lunar regolith produce a variety of secondary particles, including charged pions ($\pi^\pm$) and charged kaons ($K^\pm$). Most of these mesons either interact or decay within the regolith. Nevertheless, a fraction of the secondary particles escape the lunar surface, contributing to outgoing fluxes of muons and other particles. Lorentz-boosted muons are particularly interesting because their long lifetimes allow them to travel substantial distances through space.

We use FLUKA~\cite{Fluka,Ahdida:2022gjl,Battistoni:2015epi} to simulate the production and propagation of secondary cosmic-ray muons in lunar regolith and surrounding space. The simulation employs a simplified three-layer geometry. The Moon is modeled as a perfect sphere with a radius of $R_{\mathrm{m}} = 1737.1~\mathrm{km}$, neglecting surface topography and deviations from spherical symmetry. The lunar body is surrounded by a spherical vacuum region, enclosed by a FLUKA blackhole layer that absorbs particles leaving the simulation volume.

The nominal lunar regolith is modeled using the material composition and density profiles of the Moskalenko--Porter 2007 (MP2007) model~\cite{Moskalenko2007}, providing a realistic description of the lunar subsurface environment. An isotropic galactic cosmic-ray source is initialized on a spherical surface located $0.05~\mathrm{m}$ above the lunar surface, with incident particles restricted to the $2\pi$ solid angle directed toward the lunar surface. Shielding by the Earth is neglected in this conceptual design. The primary cosmic-ray flux consists of protons, alpha particles, and heavier nuclei sampled according to measured energy spectra.

Hadronic interactions are modeled using the DPMJET~\cite{Roesler:2000he,Fedynitch:2015kcn} and RQMD~\cite{Sorge:1989dy,Ballarini:2007prd} event generators implemented in FLUKA. Built-in solar modulation potentials of 465 MV and 1440 MV represent solar minimum and maximum, respectively. The corresponding galactic cosmic-ray spectra are based on the Badhwar--O'Neill model~\cite{badhwar1996galactic}, constrained by AMS~\cite{alcaraz2000cosmic,alcaraz2000helium} and BESS~\cite{sanuki2000precise} measurements, and tuned to the ICRC2001 fit~\cite{gaisser2001primary}.

The simulation program outputs differential yields with MC statistical uncertainties, normalized per incident primary particle, together with a normalization factor that converts them into differential fluxes per unit time and lunar-surface area. This factor is determined by the incident primary cosmic-ray flux and solar modulation.

\section{Results and discussion}

We study the integrated, energy-differential, and angular-differential fluxes. Error bars on the simulated fluxes show MC statistical uncertainties only. To conservatively mimic realistic measurement statistics, we simulate $10^6$ primary cosmic-ray events for each spectrum. We assume an overall detection efficiency of $95\%$ for all particle types. The simulated statistics correspond to exposure times of 90~s at our modeled solar minimum and 312~s at solar maximum for a $1~\mathrm{m}^2$ detector, based on primary fluxes of $1.17~\mathrm{cm}^{-2}\,\mathrm{s}^{-1}$ and $0.337~\mathrm{cm}^{-2}\,\mathrm{s}^{-1}$, respectively.

Owing to the rotational symmetry about the $z$-axis, the angular distribution is expressed as $dN/d\cos\theta$, where $\theta$ denotes the angle between the particle momentum and the detector surface normal, aligned with the local lunar radial direction. Following the convention adopted in Refs.~\cite{Fermi-LAT:2016tkg,Loparco:2017isq}, we present the energy spectra weighted by $E^2 = E_\mathrm{low}E_\mathrm{high}$, where $E_\mathrm{low}$ and $E_\mathrm{high}$ are the lower and upper boundaries of each energy bin.

We validate the nominal regolith model and event normalization by comparing the simulated lunar secondary photon flux with \textit{Fermi}-LAT measurements~\cite{Fermi-LAT:2016tkg}. Details are provided in Appendix~\ref{app:fluxes-photon}.

To characterize lunar secondary muons, we first investigate the differential muon fluxes at the lunar surface and at a representative altitude of 1 km, where meson decays are essentially complete while muon decays begin to become significant. We then examine meson production and backgrounds before assessing tomography with and without overburden.

\subsection{Muon flux at the surface}

\begin{figure}
\centering
\includegraphics[width=\linewidth]{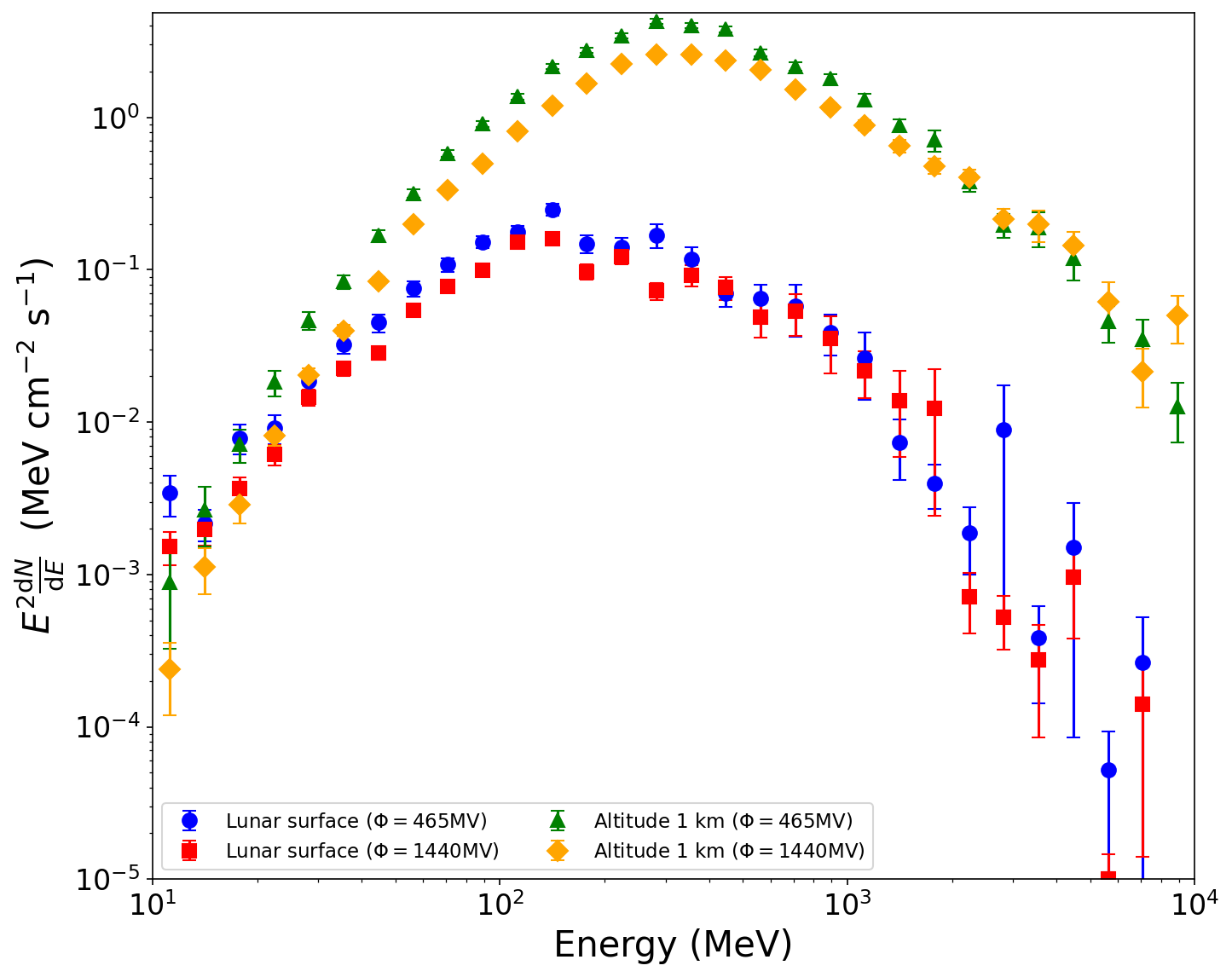}
\caption{Simulated energy-differential muon fluxes at different altitudes for various levels of solar modulation potential $\Phi$, weighted by $E^2$. Each spectrum is obtained from $N=10^6$ primary cosmic-ray events.}
\label{fig:fluxes-E}
\end{figure}

We simulate the muon flux at the lunar surface and present the energy-differential results in Fig.~\ref{fig:fluxes-E}. The spectrum exhibits a broad peak at \(E \sim 10^2~\mathrm{MeV}\), where \(E^2 \frac{dN}{dE}\) exceeds \(10^{-1}~\mathrm{MeV\,cm^{-2}\,s^{-1}}\). For energies above this peak, the spectrum follows an approximate power law, \(E^2 \frac{dN}{dE} \propto E^{-2}\).

The integrated muon flux for kinetic energies above $100~\mathrm{MeV}$ is $N(E_\mathrm{k}>100~\mathrm{MeV}) = (7.12 \pm 0.33)\times10^{-4}~\mathrm{cm}^{-2}\mathrm{s}^{-1}$ for a solar modulation potential of $465~\mathrm{MV}$, and $N(E_\mathrm{k}>100~\mathrm{MeV}) = (5.19 \pm 0.19)\times10^{-4}~\mathrm{cm}^{-2}\mathrm{s}^{-1}$ for $1440~\mathrm{MV}$. For comparison, the integrated muon flux at sea level on Earth is approximately $1.7\times10^{-2}~\mathrm{cm}^{-2}\mathrm{s}^{-1}$. Although the predicted lunar muon flux is approximately an order of magnitude lower than that at Earth's surface, it remains non-negligible. This reduction is primarily because many charged mesons escaping the lunar surface have insufficient path length to decay before reaching the detector.

\subsection{Muon flux at a height}

We also simulate the muon flux at $1~\mathrm{km}$ altitude, shown in Fig.~\ref{fig:fluxes-E} for comparison. The integrated muon flux for kinetic energies $E_\mathrm{k} > 100~\mathrm{MeV}$ is $N(E_\mathrm{k}>100~\mathrm{MeV}) = (1.22 \pm 0.02)\times10^{-2}~\mathrm{cm}^{-2}\mathrm{s}^{-1}$ for a solar modulation potential of $465~\mathrm{MV}$, and $N(E_\mathrm{k}>100~\mathrm{MeV}) = (7.59 \pm 0.08)\times10^{-3}~\mathrm{cm}^{-2}\mathrm{s}^{-1}$ for $1440~\mathrm{MV}$. At this altitude, the lunar muon flux is of the same order of magnitude as the muon flux at Earth's sea level.

Compared with the lunar surface, the muon flux at $1~\mathrm{km}$ altitude is generally more than an order of magnitude larger. This enhancement arises because the observed muons are produced primarily through the decay of charged pions, which have a proper lifetime of $\tau = 2.6 \times 10^{-8}~\mathrm{s}$~\cite{ParticleDataGroup2024}, corresponding to a proper decay length of $c\tau = 7.8~\mathrm{m}$. As a result, many charged mesons reach the lunar surface before decaying, but can subsequently decay while propagating through the near-vacuum above, producing significantly more muons that can be detected at higher altitudes such as $1~\mathrm{km}$.

On the other hand, the differential enhancement of the flux at $1~\mathrm{km}$ relative to that at the lunar surface increases from about 5 at $E \sim 10^2~\mathrm{MeV}$ to more than 100 for energies above $1~\mathrm{GeV}$. This trend arises because higher-energy muons experience larger Lorentz boosts, making them less likely to decay before reaching the detector. Consequently, the spectral peak is also shifted slightly toward higher energies, reflecting the energy-dependent survival probability of muons.

\subsection{Fluxes of other lunar secondary cosmic rays}

\begin{figure}
\centering
\includegraphics[width=\linewidth]{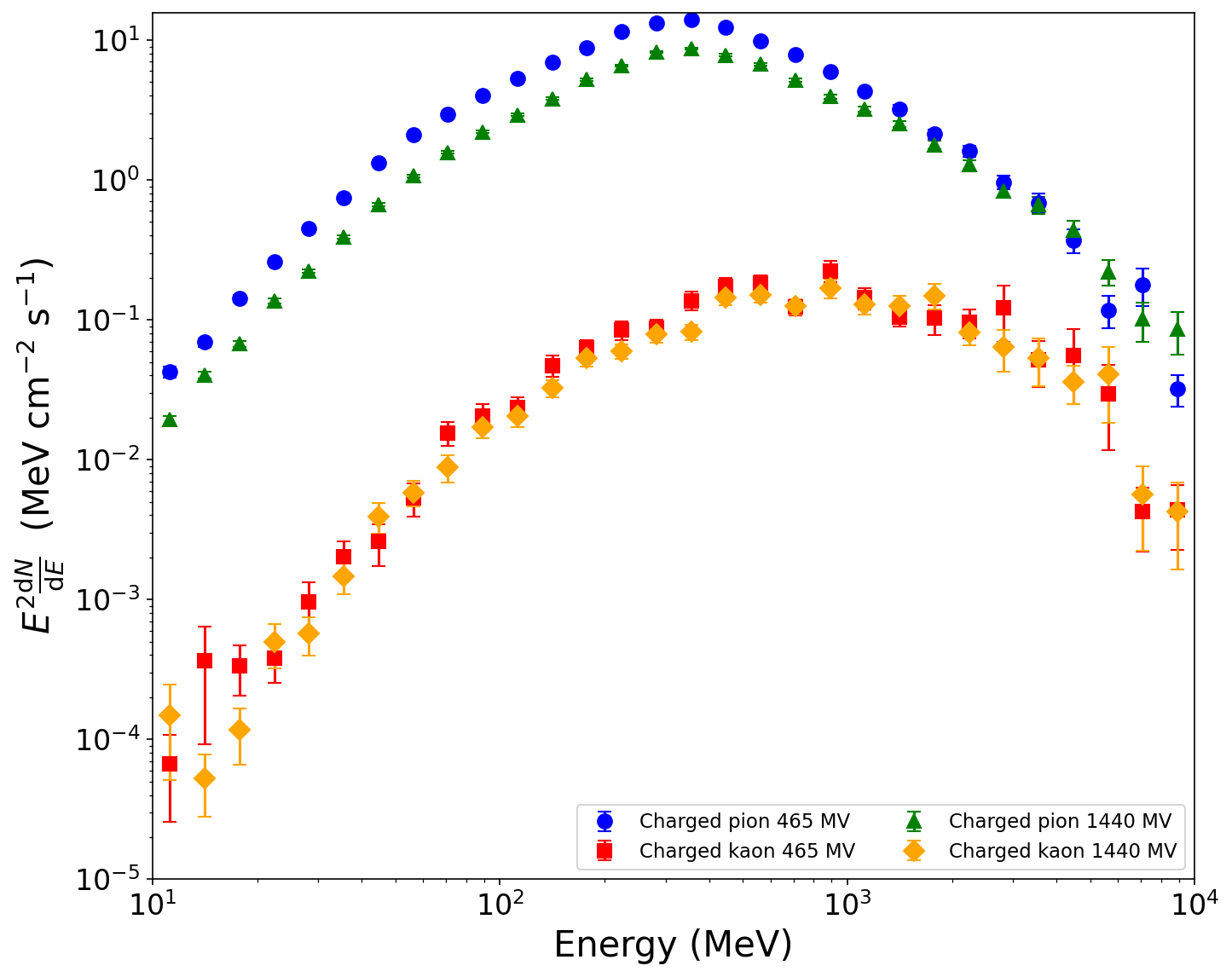}
\caption{Simulated energy-differential fluxes of charged pions and charged kaons at the lunar surface, for various levels of solar modulation potential $\Phi$. Each distribution is obtained from $N=10^6$ primary cosmic-ray events.}
\label{fig:fluxes-pion-kaon}
\end{figure}

Fig.~\ref{fig:fluxes-pion-kaon} shows the energy distributions of $\pi^\pm$ and $K^\pm$ at the lunar surface. The statistics correspond to upward-going mesons that can subsequently decay into muons. Charged pions, whose flux is about two orders of magnitude larger than that of charged kaons, dominate secondary muon production on the Moon. Together with the muon spectra in Fig.~\ref{fig:fluxes-E}, these results support displaced meson decays as the dominant source of upward-going muons.

To characterize lunar secondary backgrounds, we perform a Geant4~\cite{Allison:2016,Allison:2006,Agostinelli:2002hh} simulation, allowing straightforward extraction of individual particle species. Over $E_\mathrm{k}\sim100~\mathrm{MeV}$ to $1~\mathrm{GeV}$ at the lunar surface, upward-going neutron, photon, proton, and pion fluxes exceed the muon flux by factors of $\sim100$, while the electron flux exceeds it by $\sim10$. At low energies ($\sim100~\mathrm{MeV}$), muons can be distinguished from charged secondaries using time-of-flight (TOF) measurements. Comparing pions and muons at equal kinetic energy of $\sim100~\mathrm{MeV}$, and assuming Gaussian timing uncertainties, $90\%$ muon efficiency, and $0.1\%$ background misidentification, a $1~\mathrm{m}$ TOF system requires a resolution of $\sim49~\mathrm{ps}$ for the most challenging pion–muon separation. This could limit the total secondary background yield to $\sim10\%$ of the muon signal near $100~\mathrm{MeV}$. Compact PANDA prototypes have demonstrated $\sim30~\mathrm{ps}$ resolution per scintillator tile~\cite{Gruber:2015moa}, compatible with this two-plane TOF requirement.

At higher energies, within deep-space payload constraints, dual-readout calorimetry~\cite{Akchurin:2004fe} offers a promising approach. Simultaneous scintillation and Cherenkov measurements provide precise energy measurements and particle identification (PID). For the challenging $\mu^\pm/\pi^\pm$ separation, the combined responses offer powerful discrimination~\cite{Akchurin:2013yaa,gatto2016redtop,Chen:2025ppt}. Similar applications have been proposed for the REDTOP experiment~\cite{gatto2016redtop} and the Huizhou Hadron Spectrometer~\cite{Chen:2025ppt,An2026Huizhou}. At a muon momentum of $200~\mathrm{MeV}$, simulations suggest a pion misidentification rate of $\sim1\%$ at $90\%$ muon efficiency~\cite{Chen:2025ppt}, limiting the total secondary background yield to the same order of magnitude as the muon signal.

Downward-going galactic cosmic-ray particles, despite their flux being approximately three orders of magnitude higher than that of lunar secondary muons, can be rejected much more effectively than upward-going secondary background particles using a TOF system or a timing-based coincidence trigger. For example, the AMS-01 TOF system reported a wrong-direction misidentification probability below $10^{-8}$ with a timing resolution of approximately $120~\mathrm{ps}$, while maintaining a trigger efficiency above $99.9\%$~\cite{Alvisi:1999tof,Baldini:2001xm}. Such a high galactic background rejection efficiency could render their contribution negligible in the measured upward-going muon spectra, with minimal loss of muon signal.

\subsection{Lunar subsurface tomography}

\begin{figure}
\centering
\includegraphics[width=\linewidth]{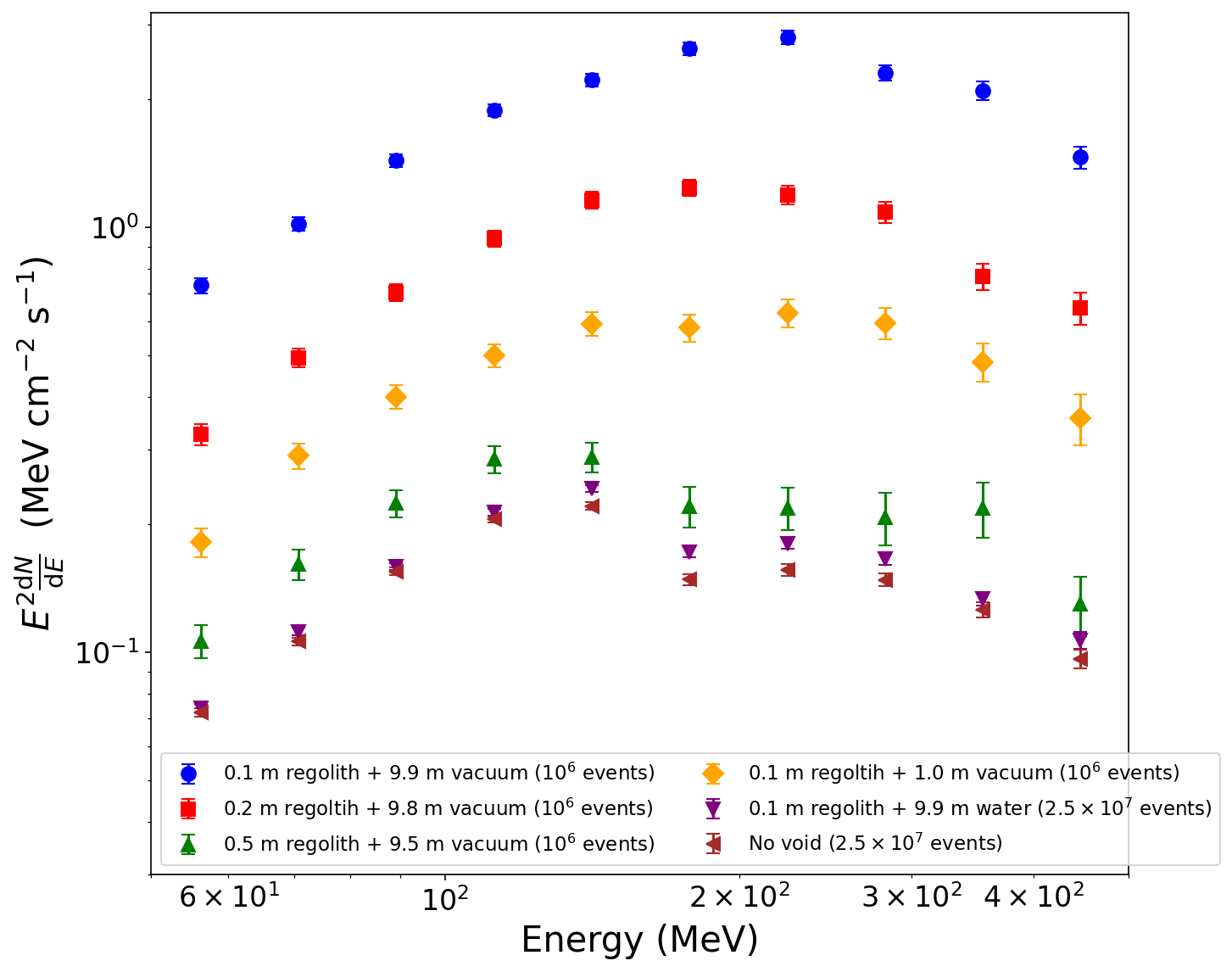}
\caption{Simulated energy-differential muon flux at the lunar surface under a solar modulation potential of $\Phi=465~\mathrm{MV}$. The curves correspond to different assumed subsurface structures beneath the lunar surface. The number of simulated primary cosmic-ray events for each distribution is indicated in the legend. The displayed $x$-axis range is chosen to highlight the differences among the scenarios.}
\label{fig:subsurface}
\end{figure}

Surface and orbital detectors motivate two distinct tomography proposals:
\begin{itemize}[leftmargin=*,itemsep=0em]
\item Lunar microscopic tomography: using detectors deployed on the lunar surface to probe subsurface structures at depths of up to $\sim 1~\mathrm{m}$, with sensitivity arising from displaced charged-meson decays into muons.
\item Lunar macroscopic tomography: using detectors in lunar orbit to probe large-area deep voids located at or near the lunar surface, with sensitivity arising from displaced muon decays.
\end{itemize}
Here, voids broadly refer to large-scale subsurface material heterogeneities, including near-vacuum cavities as well as underground water or ice deposits.

Appendix~\ref{app:fluxes-muon} illustrates the physical origin of both tomography techniques using simulated altitude-dependent muon fluxes under idealized conditions. In addition, Appendix~\ref{app:angular-muon} shows that the angular asymmetry of the muon flux at an altitude of $1~\mathrm{km}$ also reflects differences in muon propagation distance and decay probability before reaching the detector, potentially providing an additional handle for improving the sensitivity of lunar muon tomography.

We next assess the sensitivity to subsurface cavities and water deposits beneath a regolith overburden as two case studies. We model shallow subsurface anomalies as extending vertically from selected depths ($0.1$, $0.2$, and $0.5~\mathrm{m}$), above which the nominal lunar regolith is retained, to a fixed depth of $10~\mathrm{m}$. The anomalies are assumed to extend laterally over an area at least as large as the detector. This terminating depth represents the level at which the tomography sensitivity for ordinary regolith or deposited water effectively vanishes, while also exceeding the proper decay length of charged pions, allowing most charged pions escaping from the bottom of a near-vacuum cavity to decay before further interactions.

To assess the scale of systematic effects, we consider the following estimates. For solar modulation, we estimate a $10\%$ spectral-shape uncertainty over $E_\mathrm{k}=50$--$500~\mathrm{MeV}$ from the maximum deviation of the two solar-modulation spectra from their average, treated as $1\,\sigma$. For hadronic-interaction modeling, we consider an illustrative $20\%$ uncertainty, uncorrelated between energy bins, motivated by the cross-section uncertainties discussed in Ref.~\cite{Ballarini:2024fluka}. Detrended Chang'e-4 terrain data~\cite{Guo:2021roughness} in $1~\mathrm{m^2}$ patches yield a local roughness estimate of $\sigma_h=0.88~\mathrm{cm}$, corresponding to an estimated $1.8\%$ flux variation from the near-surface logarithmic derivative of $1.96~\mathrm{m}^{-1}$ (Appendix~\ref{app:semi-quantitative}).

As shown in Fig.~\ref{fig:subsurface}, shallow cavities produce several-fold to order-of-magnitude enhancements in the surface muon spectrum. These large changes remain detectable in the presence of the estimated backgrounds and systematic uncertainties. The cavity spectra correspond to $90~\mathrm{s}$ with a $1~\mathrm{m^2}$ detector assuming $95\%$ detection efficiency. Reducing the cavity's vertical extent to $1~\mathrm{m}$ reduces the enhancement, which nevertheless remains substantial. Cavity-induced changes remain visible for overburden thicknesses up to $0.5~\mathrm{m}$, although the smaller enhancements require tighter control of backgrounds and systematics. The weaker water-induced modulations at decimeter-scale depths reach $5\,\sigma$ after approximately $38$ minutes (25 times longer) only when statistical uncertainties are considered. Their detection is challenging once backgrounds and systematic uncertainties are included, and this exposure does not establish a realistic water-detection time.

\section{Summary and outlook}

In this work, we propose lunar subsurface tomography based on secondary cosmic-ray muons produced by interactions between galactic cosmic rays and the lunar regolith. Using a FLUKA-based MC framework, we simulate the production and propagation of lunar secondary particles and study the energy spectra, angular distributions, and fluxes of escaping muons for detectors on the lunar surface and in near-lunar orbit. We find that, despite the lack of a lunar atmosphere, a non-negligible muon flux emerges from the surface, with strong altitude dependence arising from displaced meson and muon decays in vacuum. These features motivate microscopic and macroscopic muon tomography studies for probing subsurface structures and resource deposits. Our case studies show several-fold to order-of-magnitude cavity-induced flux enhancements for shallow overburden. These remain detectable with the estimated backgrounds and systematic uncertainties. Water detection is more challenging because the weaker spectral modulations are more susceptible to backgrounds and systematic uncertainties. These results highlight the potential of lunar muon tomography for non-invasive subsurface characterization and resource prospecting, motivating further studies with realistic topography, detector designs, and reconstruction methods.

The concept is timely for upcoming lunar missions. Chang'e 7~\cite{wang2024scientific}, planned for launch around 2026, will explore the lunar south polar region, while Chang'e 8~\cite{CNSA_Change8_2025}, planned around 2029, will conduct scientific exploration and in-situ resource utilization experiments in support of the future International Lunar Research Station.

\begin{acknowledgments}
This work is supported in part by the National Natural Science Foundation of China (Grant No.~12325504) and the Fundamental and Interdisciplinary Disciplines Breakthrough Plan of the Ministry of Education of China (Grant No.~JYB2025XDXM204). ChatGPT and Gemini generated the background of Fig.~\ref{fig:muon-moon-shot}, with substantial improvements by Youpeng Wu. We thank Zhangshengyu Gu for contributions to simulation validation.
\end{acknowledgments}


\bibliography{apssamp}

\appendix

\section{Simulated photon flux as a validation benchmark}\label{app:fluxes-photon}

\begin{figure}
\centering
\includegraphics[width=\linewidth]{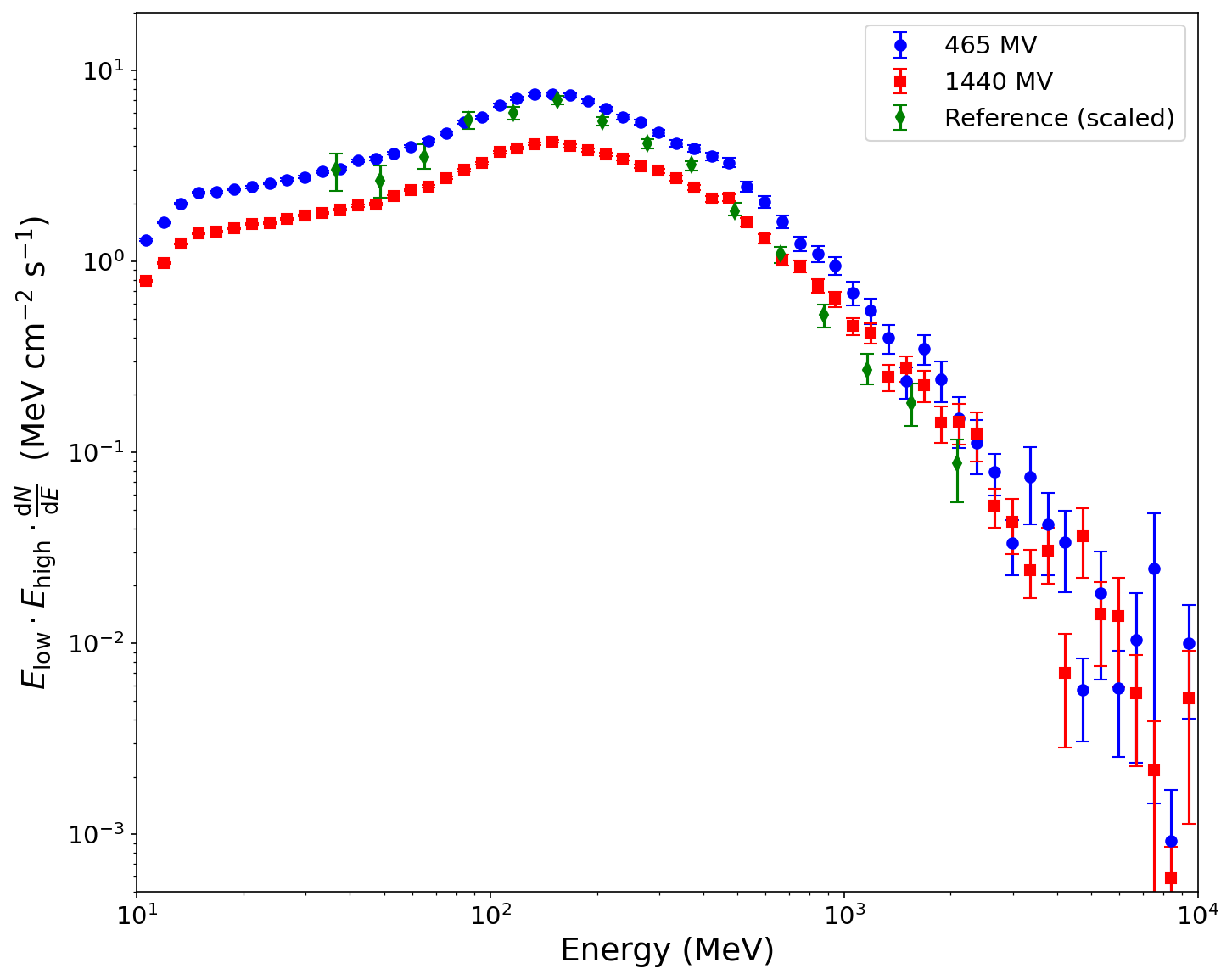}
\caption{Simulated energy-differential photon fluxes at the lunar surface for various levels of solar modulation potential $\Phi$, weighted by $E^2$. Each spectrum is obtained from $N=10^6$ primary cosmic-ray events. The data collected by the \textit{Fermi} satellite~\cite{Fermi-LAT:2016tkg} are also shown for comparison.}
\label{fig:fluxes-photon}
\end{figure}

We use the simulated lunar secondary photon flux to validate our nominal lunar regolith modeling and event normalization. In the regolith, photons are produced primarily via neutral pion decays, whose very short lifetime ($\sim 10^{-16}~\mathrm{s}$) ensures prompt decay before significant energy loss in the surrounding material. In the absence of a substantial lunar atmosphere, there is no overlying gas to cause attenuation, scattering, or altitude dependence in the secondary cascade. Consequently, the photon flux is nearly independent of observational altitude, and we therefore examine only the photon spectra at the lunar surface in the following.

Fig.~\ref{fig:fluxes-photon} shows the resulting energy-differential photon spectra at the lunar surface for various solar modulation potentials. As expected, the spectral shape changes only slightly between typical solar minimum and maximum modulation conditions, while the overall flux normalization is suppressed by approximately a factor of two. Since the photon flux is independent of the observational altitude, we compare the simulated spectra directly with the \textit{Fermi}-LAT measurements~\cite{Fermi-LAT:2016tkg}. The comparison provides validation of the normalization and spectral shape within the simulated energy range, particularly below the GeV scale, where the statistical precision is highest.

\section{Simulated muon fluxes at various altitudes}\label{app:fluxes-muon}

\begin{figure}
\centering
\includegraphics[width=\linewidth]{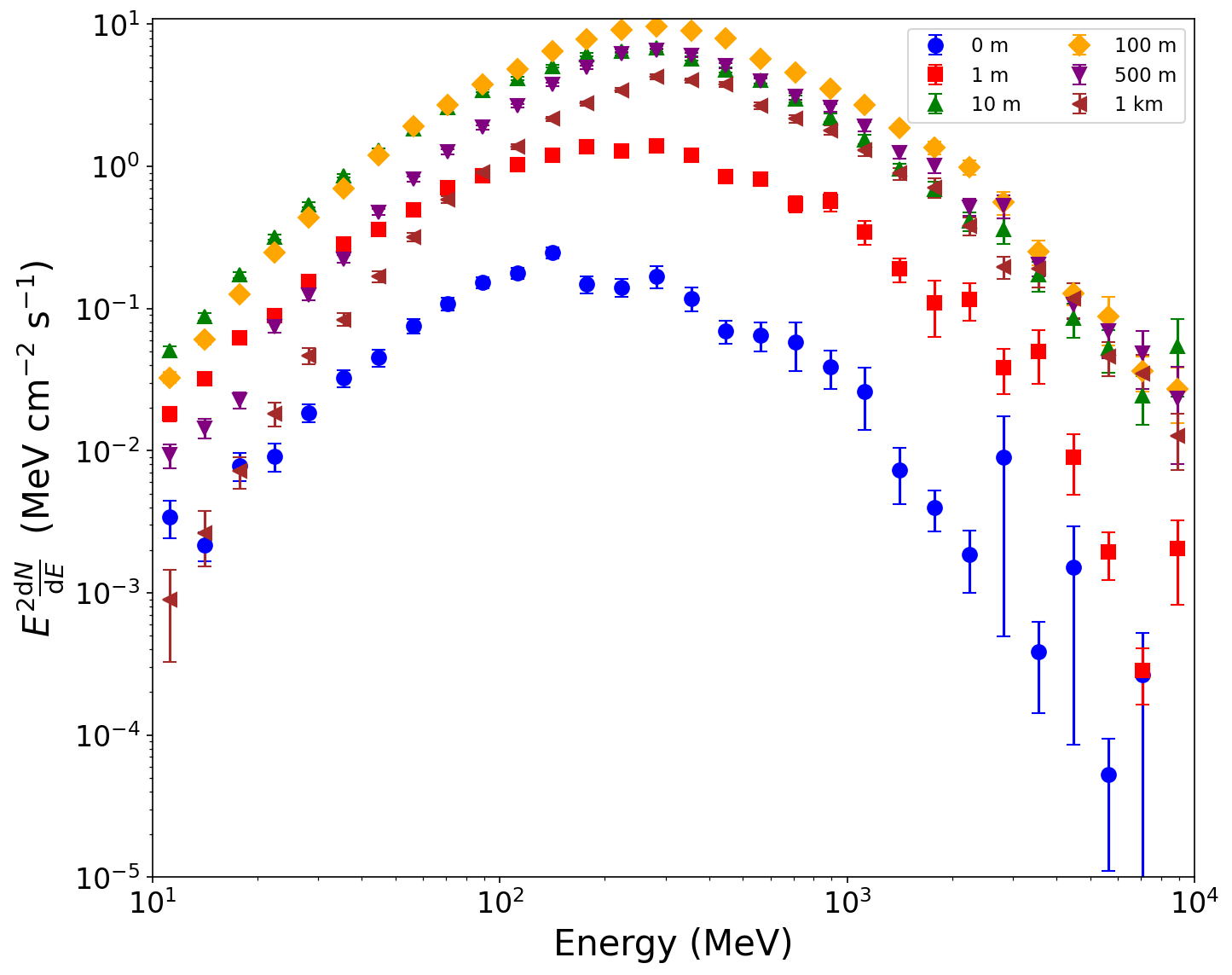}
\caption{Simulated energy-differential muon fluxes at various altitudes, for solar modulation potential $\Phi = 465~\mathrm{MV}$, weighted by $E^2$. Each distribution is obtained from $N=10^6$ primary cosmic-ray events.}
\label{fig:fluxes-muon}
\end{figure}

Due to the minor impact of solar modulation on the shape of the lunar energy-differential muon fluxes, complementary to Fig.~\ref{fig:fluxes-E}, we focus on \(\Phi = 465~\mathrm{MV}\) and show the distributions at finely sampled altitudes in Fig.~\ref{fig:fluxes-muon}. Since the muon fluxes at all considered altitudes yield ample statistics with a $1~\mathrm{m^2}$-scale detector within only several seconds, data accumulation time is not a limiting factor. Based on these distributions, we demonstrate the feasibility of muon tomography in hypothetical scenarios with a flat surface and a single type of vacuum void located sufficiently close to the surface, such that the overlying material produces negligible cosmic-ray showering effects. Although this scenario is highly idealized and not directly applicable in practice, it is sufficient to illustrate the origin of tomographic sensitivity.

For lunar microscopic tomography, we consider a scenario in which a vacuum void is located $1~\mathrm{m}$ below the lunar surface. This overlying distance allows a significant fraction of charged mesons to decay into muons, resulting in an approximately one-order-of-magnitude enhancement in the muon flux. This enhancement corresponds to the increase in effective altitude from $0$ to $1~\mathrm{m}$, as shown in Fig.~\ref{fig:fluxes-muon}. Similar tomographic sensitivity can be achieved for voids located at greater depths; however, the attenuation of primary cosmic rays while traversing the overlying material gradually offsets the muon-flux enhancement and eventually reduces the sensitivity.

In contrast, for lunar macroscopic tomography, we consider a scenario in which a large vacuum void extends from near the lunar surface to a depth of $500~\mathrm{m}$. If the muon flux is measured at an altitude of $500~\mathrm{m}$, the region above the void effectively corresponds to an altitude of $1000~\mathrm{m}$. As shown in Fig.~\ref{fig:fluxes-muon}, this increase in effective altitude from $500~\mathrm{m}$ to $1000~\mathrm{m}$ leads to an approximately one-order-of-magnitude reduction in the muon flux below $100~\mathrm{MeV}$, as the additional propagation distance allows more muons to decay before reaching the detector. However, if the void extends to greater depths, the increased production of muons in the additional overlying material gradually compensates for this reduction, thereby also diminishing the tomographic sensitivity.

To directly demonstrate that the sensitivity originates from two sequential decay processes, namely charged-meson decays followed by muon decays, we independently simulate these processes using Geant4~\cite{Allison:2016,Allison:2006,Agostinelli:2002hh}. We use the energy- and angular-differential distributions at the lunar surface as input and propagate the particles to all altitudes considered in this work. The Geant4 results show exact agreement with FLUKA. While deriving a closed-form solution for these coupled decay processes in three-dimensional spherical geometry is impractically complex and beyond the scope of this work, we provide a simplified semi-quantitative formulation in Appendix~\ref{app:semi-quantitative} to offer a clear and intuitive picture of the underlying physics.

\section{Angular distributions of lunar muons}\label{app:angular-muon}

\begin{figure}
\centering
\includegraphics[width=\linewidth]{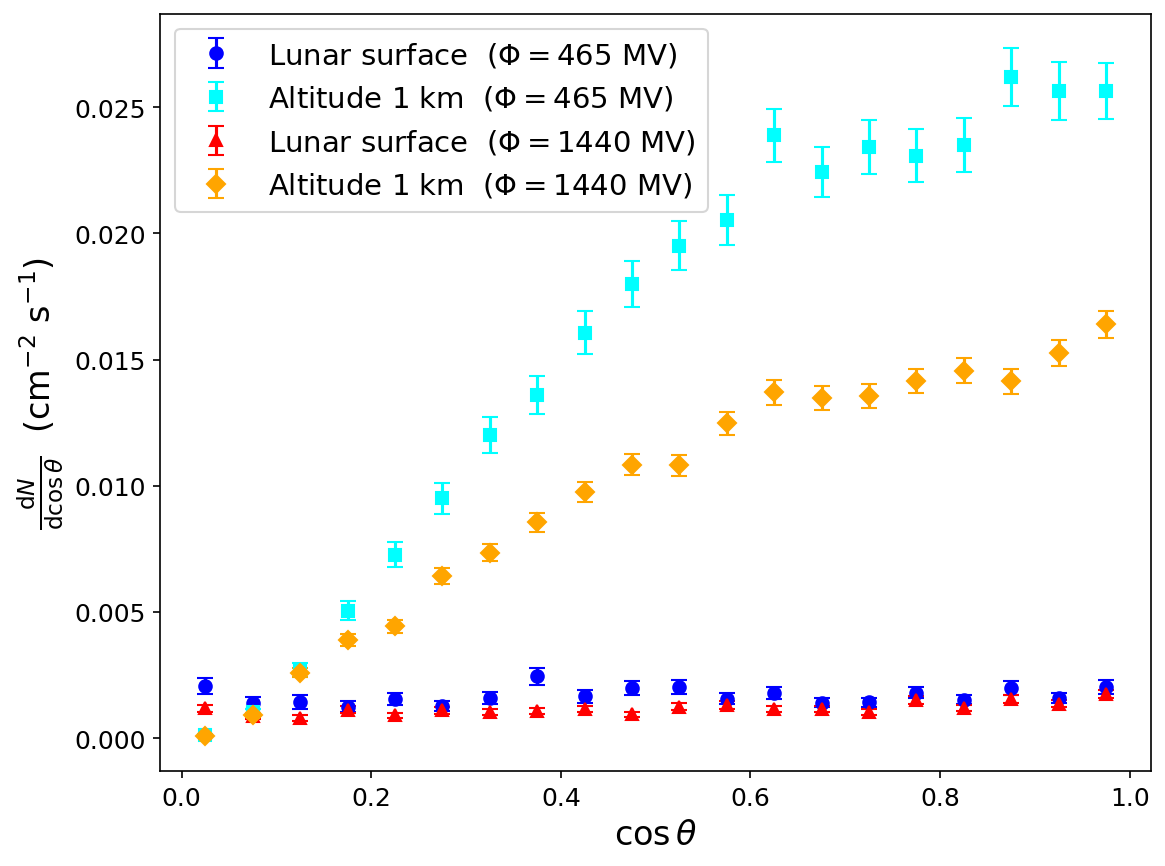}
\caption{Simulated differential angular distributions at different altitudes, for various levels of solar modulation potential $\Phi$. Each distribution is obtained from $N=10^6$ primary cosmic-ray events.}
\label{fig:fluxes-costheta}
\end{figure}

Fig.~\ref{fig:fluxes-costheta} shows the simulated differential angular distributions at different altitudes for various levels of solar modulation. In addition to the increased flux, the distribution at the lunar surface is approximately isotropic, whereas at an altitude of $h = 1~\mathrm{km}$ the flux at near-horizontal angles is significantly suppressed. This behavior is primarily attributed to muon decay: muons arriving at the detector from near-horizontal directions traverse substantially longer path lengths than vertically incident muons, increasing their probability of decaying before reaching the detector.

Quantitatively, with a proper lifetime of $\tau = 2.2 \times 10^{-6}~\mathrm{s}$, a muon with a kinetic energy of $100~\mathrm{MeV}$ has a mean decay length of $\lambda_{\mu,100\,\mathrm{MeV}} = 1.1~\mathrm{km}$. A vertically incident muon traverses approximately 1 km of vacuum, corresponding to a survival probability of roughly $40\%$. By contrast, a muon traveling along a near-horizontal trajectory must traverse a chord of length $l = \sqrt{(R_{\mathrm m}+h)^2-R_{\mathrm m}^2} \approx 59~\mathrm{km} \approx 54\lambda_{\mu,100\,\mathrm{MeV}}$ before reaching the detector at altitude $h$. Such a long flight path leads to substantial decay losses, strongly suppressing the near-horizontal muon flux, particularly at low energies.

The physical origin of the angular distributions suggests that they effectively encode information about the propagation distance, thereby providing a potential additional handle for a more precise determination of the effective measurement altitude. Exploiting this angular dependence may enhance the sensitivity of lunar muon tomography by effectively probing a range of measurement altitudes beyond that accessible through the total flux alone. We leave a quantitative investigation of this potential to future studies incorporating realistic detector simulations and dedicated reconstruction frameworks.

\section{Semi-quantitative model of muon flux dependence on decay distance}\label{app:semi-quantitative}

To illustrate the physical origin of the strong altitude dependence of the lunar muon flux, we develop a simplified semi-quantitative model of the coupled charged-meson and muon decay processes. Since a full treatment of the three-dimensional geometry and angular distributions is analytically impractical, we adopt a one-dimensional approximation in which altitude is replaced by particle propagation distance. This simplification retains the essential decay physics and provides an intuitive understanding of the sensitivity observed in the simulations.

We consider the initial fluxes of muons, charged pions, and kaons at $h=0$, denoted by $F_{\mu}(E,0)$, $F_{\pi}(E,0)$, and $F_{K}(E,0)$, respectively, where $h$ denotes the propagation distance along the single spatial axis of the model. The fraction of muons emitted in the forward direction from charged-meson decays is described by an energy-dependent factor $g(E)$, ranging from 0.5 to 1 depending on the relativistic boost. Since the initial kaon flux is one to three orders of magnitude lower than the pion flux, we neglect its contribution in this semi-quantitative approximation. Integrating the coupled decay processes, the muon flux after a propagation distance $h$ can be expressed as $F_{\mu}(E,h)=F_{\mu}(E,0)e^{-\frac{h}{\gamma_{\mu}\beta_{\mu}c\tau_{\mu}}}+F_{\pi}(E',0)g(E)\operatorname{Br}(\pi \rightarrow \mu+\nu)\frac{dE'}{dE}\frac{\gamma_\mu\beta_\mu\tau_\mu}{\gamma_\mu\beta_\mu\tau_\mu-\gamma_\pi\beta_\pi\tau_\pi}(e^{-\frac{h}{\gamma_\mu\beta_\mu c\tau_\mu}}-e^{-\frac{h}{\gamma_\pi\beta_\pi c\tau_\pi}})$. Here, $E$ and $E'$ denote the kinetic energies of muons and their parent pions, respectively, while $\gamma$ and $\beta$ are the corresponding Lorentz factors and speeds normalized to the speed of light. The first term describes the survival of initially produced muons, while the second accounts for muons produced through pion decays during propagation.

To quantify the sensitivity to propagation distance, we evaluate the logarithmic derivative of the muon flux near $h=0$: $\frac{d\ln F}{dh}\approx\frac{F_\pi(E')g\operatorname{Br}dE'}{F_\mu(E)\gamma_\pi\beta_\pi c\tau_\pi dE}$. From the simulated surface spectra, the pion-to-muon flux ratio reaches approximately 20 in the most sensitive energy region. Taking $g=1$, $\operatorname{Br}=1$, and $\frac{dE'}{dE}\approx1.75$, we estimate a logarithmic sensitivity of $1.96~\mathrm{m}^{-1}$ at a muon kinetic energy of $\sim100~\mathrm{MeV}$. This indicates a strong flux enhancement over a propagation distance of approximately one meter, consistent with the order-of-magnitude increase observed in Fig.~\ref{fig:fluxes-muon}.

\end{document}